
\def\setup{\count90=0 \count80=0 \count91=0 \count85=0
\countdef\refno=80
\countdef\secno=85}
\def\R{\vrule height5.85pt depth.2pt \kern-.05pt \tt R}
\def\O{\hat}
\def\G{{\cal G}}
\def\Box#1{\vcenter{\hrule
                    \hbox{\vrule height#1pt \kern#1pt \vrule height #1pt}
                    \hrule}}
\def\lo{\raise2pt\hbox{$<$}\kern-7pt\raise-2pt\hbox{$\sim$}}
\def\go{\raise2pt\hbox{$>$}\kern-7pt\raise-2pt\hbox{$\sim$}}
\def\parallel#1#2{\hbox{\kern1pt \vrule height#1pt
                                 \kern#2pt
                                 \vrule height#1pt \kern1pt}}

\def\Vline{{\vrule height13pt depth8pt}\; }
\def\sub#1{{\lower 8pt \hbox{$#1$}}}
\def\Basis{\hat\Phi}

\def\Del{{\raise.5ex\hbox{$\bigtriangledown$}}}
\def\DEL#1{{\raise.5ex\hbox{$\bigtriangledown$}\raise 8pt \hbox{\kern -10pt
                 \hbox{$#1$}} }}
\def\autoeq{ {\global\advance\count90 by1} \eqno(\the\count90) }
\def\autoeql{ {\global\advance\count90 by1} & (\the\count90) }
\def\ceist{ {\global\advance\count91 by1} (\the\count91) }
\def\autosec{ {\global\advance\secno by 1} (\the\secno) }

\def\Lie#1{{\cal L}{\kern -6pt
            \hbox{\raise 1pt\hbox{-}}\kern 1pt} _{\vec{#1}}}
\def\lie#1{{\cal L}_{\vec{#1}}}
\def\Z{Z \kern-5pt \hbox{\raise 1pt\hbox{-}}\kern 1pt}


\def\autoref{ {\global\advance\refno by 1} \kern -5pt [\the\refno]\kern 2pt}

\def\readref#1{{\count100=0 \openin1=refs\loop\ifnum\count100
<#1 \advance\count100 by1 \global\read1 to \title \global\read1 to \author
\global\read1 to \pub \repeat\closein1  }}

\def\reftitle{{ \kern -3pt \vtop{ \hbox{\title} \hbox{\author\ \pub} } }}
\def\ref{{ \kern -3pt\author\ \pub \kern -3.5pt }}

\def\refanon{{ \hbox{\pub}\kern -3.5pt }}

\def\circum#1{{ \kern -3.5pt $\hat{\hbox{#1}}$ \kern -3.5pt}}

\magnification=1200
\setup
\centerline{  }
\line{\hfill \hbox{ITP-UH-06/93} }
\line{\hfill \hbox{May 1993} }
\vskip 2cm
\centerline{{\bf The Renormalisation Group Equation As An Equation}}
\centerline{{\bf For Lie Transport Of Amplitudes}
\footnote*{Work carried out with an  Alexander von Humboldt
research stipendium.}}

\vskip 1.2cm
\centerline{Brian P. Dolan}
\vskip .5cm
\centerline{\kern 10pt {\it Institut f\"ur Theoretische Physik}
\footnote{**}{On leave of absence from: Department of Mathematical
Physics, St. Patrick\rq s
College, Maynooth, Ireland.}}
\centerline{\it Universit\"at Hannover}
\centerline{\it Hannover, Germany}
\centerline{e-mail: dolan@kastor.itp.uni-hannover.de}
\vskip 1.5cm
\centerline{ABSTRACT}
It is shown that the renormalisation group (RG) equation
can be viewed as an equation for Lie transport of
physical
amplitudes along the integral curves generated by the $\beta$-functions
of a quantum field theory. The anomalous dimensions arise from
Lie transport of basis vectors on the space of couplings.
The RG equation can be interpreted as relating a particular diffeomorphism
of flat space-(time), that of dilations, to a diffeomorphism
on the space of couplings generated by the vector field associated
with the $\beta$-functions.

\vskip 1cm

\vfill\eject

\vfill\eject

\bf \S1 Introduction\rm
\vskip.5cm
Recently there has been much interest in the possibility of
putting differential geometrical structures such as a metric and connection
on the space of interactions of a quantum field
theory.
In particular metrics on the space of interactions
have been considerd
in\autoref\newcount\Wallace\Wallace=\refno\autoref\newcount\Zamo\Zamo=\refno
and\autoref\newcount\Denjoe\Denjoe=\refno and the
question of  connections has been addressed
in\autoref\newcount\Zwiebach\Zwiebach=\refno.
A more primitive concept than covariant differentiation
on a manifold is that
of Lie differentiation and the purpose of this paper is to point
out that away from a fixed point, when the $\beta$-functions are
non-zero, the renormalisation group (RG) equation can be interpreted
as an equation for Lie transport along the vector field on the space
of couplings defined by the $\beta$-functions of the theory.
That the RG equation may be understood without the necessity of
introducing a connection on the space of couplings should not come as
a surprise, since the usual derivations of this equation do not
require introducing any extra structures, such as a connection, onto
the space of couplings.

The idea presented in this paper is the following.
A rescaling  in Euclidean space, ${\bf R}^D$, is a
diffeomorphism and is generated by the vector
$\vec{\bf D}=x^\mu\partial_\mu$. If the space of couplings is also
considered to be a finite dimensional, differentiable manifold, $\G$,
then physical
amplitudes can be interpreted as co-variant tensors on $\G$ and
one can view the rescaling of the couplings as a
diffeomorphism of $\G$ which is generated by the vector
given by the $\beta$-functions of the theory.
The renormalisation group equation then expresses
the fact that the change in amplitudes under
the diffeomorphism of ${\bf R}^D$ generated by $\vec{\bf D}$,
keeping the couplings fixed,
is exactly the same as the change effected by a diffeomorphism
of $\G$, generated by the vector $\vec\beta$, keeping the
spatial points
$x_i$ fixed. Alternatively,
if the RG rquation is written as a differential equation
involving $\kappa{\partial\over\partial\kappa}\vert_g$ rather
than $x^\mu{\partial\over\partial x^\mu}\vert_g$, using standard
naive scaling arguments, it becomes nothing more than the
{\it definition} of a Lie derivative on $\G$ with respect
to the vector field $\vec\beta$.
The terms in the RG equation involving anomalous dimensions
are interpreted as coming from the change in the basis for
co-vectors, $dg^a$, as we move along the RG trajectory.

Of course quantum field theory is famous for being plagued by
\lq\lq infinities" which, at least for renormalised theories,
can be \lq\lq tamed" by a regularisation procedure.
This requires the introduction of \lq\lq bare" couplings,
$g^a_o(g,\epsilon)$,
which are analytic functions of the renormalised couplings, $g^a$,
and a regularisation parameter or paramaters, $\epsilon$.
e.g.~for a cut-off, $\Lambda$, $\epsilon=\kappa/\Lambda$ where
$\kappa$ is a renormalisation point and for
dimensional regularisation $\epsilon=4-D$ where $D$ is
the dimension of space or space-time.
The point of view adopted here is that $g_o^a$ and $g^a$
can be thought of as different co-ordinate systems on $\G$.
The matrix ${\partial g_o^a\over\partial g^b}$ tells us
how to transform tensors (amplitudes) between co-ordinate systems.
$g_o^a$ enter on a different footing from $g^a$
however in that they depend on the regularisation parameter
whereas $g^a$ do not.
The \lq\lq infinities" of quantum field theory are then viewed
as being due to the fact that the co-ordinate transformation
between $g_o^a$ and $g^a$
is singular when the regularisation parameter is removed,
but this is not an insurmountable problem as long as theory is
renormalisable.
One must be very careful to distinguish between genuine
singularities and singularities that are merely due to the choice
of co-ordinates.

The basic idea presented here, that the RG equation should be
viewed in terms of a co-ordinate transformation on the
space of couplings, was inspired by O'Connor and
Stephens [\the\Denjoe].

\vskip 1cm
\noindent
{\bf \S2 The Renormalisation Group Equation As A Lie Derivative}
\vskip .5cm

It will now be demonstrated that the RG equation
can be interpreted as an equation for
Lie derivatives of amplitudes. The Lie derivative of an
amplitude with respect to the dilation generator, $\vec{\bf D}$,
on Euclidean space is exactly the same as the Lie derivative
with respect to the vector
$\vec{\beta}=\beta^a\partial_a$ on the space of couplings.

Consider a field theory in flat $D$-dimensional Euclidean space.
In principle there are
an infinite number of operators that can be constructed out of the
fields, each of which introduces a coupling,
but the criterion of renormalisability requires that only a finite
number of these couplings is
independent,\autoref\newcount\WilKog\WilKog=\refno.
This means that, within the a priori infinite dimensional space
of coupling constants, the theory can be formulated on a
finite $n$-dimensional subspace which will be denoted
by $\G$. It will be assumed
that $\G$ is, at least locally, a differentiable manifold
and we will denote the renormalised couplings (co-ordinates on $\G$)
by $g^a; a=1,\ldots ,n$.
Bare quantities will be represented by a subscript $o$. Thus
the bare couplings are denoted by $g_o^a$. It will be assumed
that a regularisation procedure is imposed which renders bare
quantities very large but still finite. This is because bare
quantities appear in some of the following formulae and we
do not want to be manipulating infinite quantities. Ultimately,
of course, all bare quantities disappear from physical amplitudes.
The couplings
$g^a$ will be taken to be real and massless.
If the theory contains any couplings which are massive
these can always be made massless
by multiplying by appropriate powers of the renormalisation
mass scale, $\kappa$.
Questions about the global structure of $\G$ will not be addressed here.

Following ref. [\the\Zamo] we consider the operators
\newcount\Basic
$$\hat \Phi_a(x)={\partial \hat H_o(x)\over\partial g^a},\autoeq$$
\Basic=\count90
where $\O H_o(x)$ is the bare Hamiltonian density,
to be a basis for all relevant
or marginal operators of the theory,
i.e. any relevant or marginal
operator, which is a scalar (or more precisely a density)
in ${\bf R}^D$, can be written
as a linear combination of $\Basis_a(x)$.

Since $\Basis_a$ are a basis for all relevant (and marginal)
rotationally invariant operators
it suffices to consider
amplitudes of
these operators. All other amplitudes for Euclidean
scalars (more precisely densities) which are relevant or
marginal operators can be obtained from linear
combinations of these basic operators. Consider, therefore,
amplitudes of the form
$$\Basis^{(N)}_{a_1\cdots a_N}(x_1,\ldots x_N)=
<\Basis_{a_1}(x_1)\cdots\Basis_{a_N}(x_N)>.\autoeq$$
Of course $\Basis_a$ are, in general, composite operators
and renormalisation of amplitudes involving multi-insertions of
composite operators is a subtle problem involving the removal
of new infinities which are not present in amplitudes involving
only the elementary fields. A clear treatment of this
problem is given in\autoref\newcount\nlsr\nlsr=\refno.
However, these problems can be avoided here since we shall always
assume that $x_i\ne x_j$ $\forall i\ne j$ in amplitudes, so these
extra infinities never arise.

The usual derivation of the renormaliation group equation
relies on the simple fact that all physical amplitudes should be
independent of the renormalisation point $\kappa$.
However $\Basis^{(N)}_{a_1\cdots a_N}(x_1,\ldots x_N)$ cannot be
independent of $\kappa$ in general. If we chose a different
parameterisation of the renormalised couplings (a different co-ordinate
system on $\G$), which will be denoted by primes $g^{a^\prime}(g)$, then
$\Basis^{(N)}_{a_1\cdots a_N}(x_1,\ldots x_N)$ transforms
as a rank $N$ co-variant tensor,
$$\Basis^{(N)}_{a_1\cdots a_N}(x_1,\ldots x_N)=
\left({\partial g^{a^\prime_1}\over\partial g^{a_1}}\right)\cdots
\left({\partial g^{a^\prime_N}\over\partial g^{a_N}}\right)
\Basis^{(N)}_{a^\prime_1\cdots a^\prime_N}(x_1,\ldots x_N),\autoeq$$
and these cannot both be independent of $\kappa$, since
${\partial g^{a^\prime_i}\over\partial g^{a_i}}$ is not, in general.
The object that ought to be independent of $\kappa$ is
the reparameterisation invariant amplitude
$$<\Basis(x_1)\cdots\Basis(x_N)>=
<\Basis_{a_1}(x_1)\cdots\Basis_{a_N}(x_N)>dg^{a_1}\ldots dg^{a_N}.
\autoeq$$
The correct statement that amplitudes
are independent of the renormalisation point is,
$$\kappa{d\over d\kappa}<\Basis(x_1)\cdots\Basis(x_N)>=0.
\autoeq$$
Allowing for the fact that $<\Basis_{a_1}(x_1)\cdots\Basis_{a_N}(x_N)>$
are also functions of the $g^a$ gives
$$\eqalign{
-\kappa{\partial\over\partial \kappa}\Vline\sub g &
<\Basis_{a_1}(x_1)\cdots\Basis_{a_N}(x_N)>=\cr
&\beta^b\partial_b<\Basis_{a_1}(x_1)\cdots\Basis_{a_N}(x_N)>
+\sum_{i=1}^N\bigl(\partial_{a_i}\beta^b\bigr)
<\Basis_{a_1}(x_1)\cdots\Basis(x_i)_{b}\cdots\Basis_{a_N}(x_N)>,\cr}
\autoeq$$
where we have used
$$\kappa{d\over d\kappa}(dg^a)=d\left(\kappa{dg^a\over d\kappa}\right)
=d\beta^a={\partial\beta^a\over\partial g^b}dg^b.\autoeq$$
In co-ordinate free notation this is
\newcount\rgq
$$-\kappa{\partial\over\partial \kappa}\Vline\sub g
<\Basis(x_1)\cdots\Basis(x_N)>=
\lie{\beta}<\Basis(x_1)\cdots\Basis(x_N)>,\autoeq\global\rgq=\count90$$
where $\lie{\beta}$ is the Lie derivative on $\G$ with respect to
the vector field, $\vec{\beta}$.
Thus the renormalisation group equation is nothing other
than the {\it definition} of a Lie derivative
provided that it is appreciated that the amplitudes are tensors
on $T^*(\G)$, and not scalars (the minus sign in (\the\rgq)
is standard in the definition of a Lie
derivative, see e.g.\autoref\newcount\HE\HE=\refno).
This analysis
makes it clear that the anomalous dimensions
have the geometrical interpretation of arising from the change
in $dg^a$ (which are a basis for real valued one-forms
on $T^*(\G)$) as we move along the vector field $\vec\beta$.

Equation (\the\rgq) can be expressed in terms of the
dilation vector on Euclidean space, $\vec{\bf D}$, using
the usual scaling arguments.
To see this first note that
$\Basis_a$ are densities ($D$-forms) in ${\bf R}^D$
therefore, in Cartesian co-ordinates,
the Lie derivative of $\Basis_a(x)$ with respect to the vector
$\vec{\bf D}$ is
$$\Lie{\bf D}\Basis_{a_i}(x_i)=
(di_{\vec{\bf D}} + i_{\vec{\bf D}}d)\Basis_a(x)=di_{\vec{\bf D}}\Basis_a(x)
={\partial\over\partial x_i^\mu}\bigl(x_i^\mu\Basis_{a_i}(x_i)\bigr)=
\Bigl(x_i^\mu{\partial\over\partial x_i^\mu} + D\Bigr)\Basis_{a_i}(x_i),
\autoeq$$
where $d$ and $i_{\vec{\bf D}}$ here represent exterior derivative
and contraction on ${\bf R}^D$.
Now by definition, equation (\the\Basic), $\Basis_a(x)$
has mass dimension $D$
thus the usual naive scaling arguments give
$$\eqalign{
\kappa{\partial\over\partial \kappa}\Vline\sub g
<\Basis_{a_1}(x_1)\cdots\Basis_{a_N}(x_N)>
&=\sum_{i=1}^N\Bigl(x_i^\mu{\partial \over\partial x_i^\mu}+D\Bigr)
<\Basis(x_1)_{a_1}\cdots\Basis_{a_N}(x_N)>\cr
&=\Lie{\bf D}<\Basis_{a_1}(x_1)\cdots\Basis_{a_N}(x_N)>
,\cr}\autoeq$$
hence
$$\Lie{\bf D}
<\Basis(x_1)\cdots\Basis(x_N)>=
-\lie{\beta}<\Basis(x_1)\cdots\Basis(x_N)>.
\autoeq\newcount\RG\global\RG=\count90$$
This is the geometrical form of the RG equation, in co-ordinate free notation.
It is seen to relate a particular diffeomorphism
(dilations generated by the conformal Killing vector, $\vec{\bf D}$)
of Euclidean space, ${\bf R}^D$,
to a diffeomorphism (generated by the associated vector $\vec{\beta}$)
of the space of couplings $\G$.

Another way of expressing this idea is to
define the matrix of renormalisation co-efficients, ${Z_a}^b$, by
$$\Basis_{oa}={Z_a}^b(g)\Basis_b,\autoeq$$
where $\Basis_{oa}$ constitute the bare basis
$$\Basis_{oa}={\partial\O H_o\over\partial g_o^a}.\autoeq$$
The matrix ${Z_a}^b$ depends on the renormalisation scheme, of course.
For example in dimensional regularisation it is a function not only
of the renormalised couplings, $g^a$, but also of the dimension
$D=4-\epsilon$ and can be expanded as a series of poles in $\epsilon$.
Clearly the definition of the renormalised basis (\the\Basic) also
implies that
$${Z_a}^b(g)={\partial g^b\over\partial g_o^a}
\qquad\qquad\Leftrightarrow\qquad\qquad
dg_o^a{Z_a}^b(g)=dg^b
\autoeq\newcount\dgz\global\dgz=\count90$$ so we
can write
$$\Basis(x)=\Basis_{oa}(x)dg_o^a=\Basis_a(x)dg^a.\autoeq$$
We now demand that the bare operators be
independent of the renormalisation point. There is a slight subtlety
here, though, in that the bare couplings are defined to be massless. This
means that they change, under changes in $\kappa$, by their
{\it canonical} dimensions. e.g. for a mass the massless bare
coupling is $\tilde m_o^2=m_o^2\kappa^{-2}$ and
$\kappa{dm_o^2\over d\kappa}=0$ requires
$\kappa{d\tilde m_o^2\over d\kappa}=-2\tilde m_o^2$.
We shall denote the canonical dimension of the coupling $g^a$ by $d_a$,
thus $\kappa{dg^a_o\over d\kappa}=-d_ag_o^a$ where there is no sum over $a$.
As in the previous argument for amplitudes
the operator-valued one form $\Basis_o=\Basis_{oa}dg_o^a$
is independent of $\kappa$. Thus
$$\kappa{d\Basis_{oa}\over d\kappa}-d_a\Basis_{oa}=0,
\qquad\qquad\Leftrightarrow\qquad\qquad
\kappa{d\Basis_a\over d\kappa}+{\Gamma_a}^b\Basis_b=0
\autoeq$$ \newcount\ORG\ORG=\count90
or
$$\left(\kappa{\partial\over d\kappa}\Vline\sub g
+ \beta^b\partial_b\right)\Basis_a+{\Gamma_a}^b\Basis_b=0,
\autoeq\newcount\exactrg\global\exactrg=\count90$$
where the matrix, ${\Gamma_a}^b$, is defined by
$${\Gamma_a}^b=
{\partial\beta^b\over\partial g^a}.\autoeq$$
This includes the matrix of anomalous dimensions,
$${\gamma^a}_b=
{\bigl(Z^{-1}\bigr)^a}_c\kappa\left({d{Z^c}_b\over d\kappa}\right),
\autoeq$$
but they are not equal because the couplings are dimensionless.
Equation (\the\exactrg) is a renormalisation group equation for the operators
$\Basis_a$. One must be careful in evaluating amplitudes
involving this equation however since neither
$\kappa{\partial\over d\kappa}\vert_g$ nor
$\beta^b\partial_b$ can be pulled outside of expectation values
separately, though the combination can be since the action
$S_o=\int d^D xH_o(x)$, which appears in the functional integral representation
of amplitudes, is
independent of $\kappa$.

It is crucial to this interpretation that massless couplings
are used. In particular this means that
$$\vec\beta=\beta^a{\partial\over\partial g^a}
=\beta_o^a{\partial\over\partial g_o^a},
\autoeq$$\newcount\trivbeta\trivbeta=\count90
where $\beta_o^a=\kappa{dg_o^a\over d\kappa}=-d_a$
are just the canonical dimensions.
Had massive couplings been used, then $\beta_o^a$ would
be zero since the bare massive couplings are defined to be independent
of $\kappa$, but $\beta^a$ are non-zreo in general.
One cannot transform from a non-zero vector to one which vanishes
using a co-ordinate transformation. (This has nothing to do
with the \lq\lq singularities" of the renormalisation program,
the regulator is still in place so the co-ordinate transformation
is still non-singular.)
The important quantity is
the total dimension, canonical plus anomalous, and to split it
up spoils general co-ordinate invariance. Stated differently,
the redefinition $m^2\rightarrow \tilde m^2=m^2\kappa^{-2}$
is {\it not} a co-ordinate transformation because $\kappa$ is
not a co-ordinate.

\vfil\eject

\bf\S 3 Conclusions \rm
\vskip .5cm

The emphasis here is on a co-variant description of the RG equation
and the notion that one is free to choose any set of renormalised
co-ordinates that one wishes in a description of physical amplitudes.
It has been argued that the renormalisation group equation
can be given a geometrical intrepretation in
the sense that it may be viewed as an equation for Lie
transport. In terms of Lie transport on the space of couplings
it reduces to no more than the definition of a Lie derivative, but
its real significance lies in the way that it ties a particular
diffeomorphism of Euclidean space, that of dilations, to the diffeomorphism
of the space of couplings generated by the vector field
$\vec\beta$ through equation (\the\RG)
$$\Lie{\bf D}
<\Basis(x_1)\cdots\Basis(x_N)>=
-\lie{\beta}<\Basis(x_1)\cdots\Basis(x_N)>,\eqno(\the\RG)$$
where $\Lie{\bf D}$ is the Lie derivative on Euclidean space
and $\lie{\beta}$ the Lie derivative on the space of couplings.
It is crucial to this interpretation that
$N$-point amplitudes
be viewed as rank $N$ co-variant tensors on the space of couplings,
and equation (\the\RG) states that the two diffeomorphisms are
completely equivalent. From this point of view the anomalous dimension
terms in the RG equation are seen as coming from the change in the
basis $dg^a$ under Lie transport along the trajectories of $\vec\beta$.
Since Lie derivatives are the natural geometric way in which to describe
symmetries, this picture gives a clearer insight into the
structure of the RG equation and the manner in which conformal
symmetry, and perhaps also other space symmetries, may be broken.
For instance one might postulate the existence of a $\beta$-function,
$\lie{\beta_X}$
associated with a more general diffeomorphism, $\vec X$, of space
and consider the equation
$$\Lie{X}
<\Basis(x_1)\cdots\Basis(x_N)>=
-\lie{\beta_X}<\Basis(x_1)\cdots\Basis(x_N)>.
\autoeq$$
This would presumably necessitate the introduction of position dependent
couplings, a concept which has already proved to be of some use in
understanding the RG
equation\autoref\newcount\gx\gx=\refno.

An alternative version of the RG equation, for operators rather
than amplitudes, is given by equation (\the\exactrg),
$$\left(\kappa{\partial\over d\kappa}\Vline\sub g
+ \beta^b\partial_b\right)\Basis_a+{\Gamma_a}^b\Basis_b=0,
\eqno(\the\exactrg)$$
but it must be stressed that neither
$\kappa{\partial\over d\kappa}\vert_ g$ nor
$\beta^b\partial_b$ can be pulled out of expectation values separately
since $S_o$ is not invariant under either separately.

It is a pleasure to thank
Denjoe O'Connor for many discussions on the geometric
nature of the RG.

After completion of this work the author's attention was drawn
to reference \autoref\newcount\lassig\lassig=\refno where the RG
equation is also discussed in terms of Lie derivatives by introducing
a connection on the space of couplings. I am grateful to Chris
Stephens for informing me of this reference.

\vfill\eject

\bf References \rm\hfill
\vskip1cm
\frenchspacing

\item{[\the\Wallace]} D.J. Wallace and R.K.P. Zia, Phys. Lett. \bf 48A\rm,
(1974), 325\hfil\break
Ann. Phys. \bf 92\rm, (1975), 142
\smallskip
\item {[\the\Zamo]}
A.B Zamolodchikov, Rev. Math. Phys. \bf 1\rm, (1990), 197
\smallskip
\item{[\the\Denjoe]} D. O'Connor and C.R. Stephens, \it
Geometry The Renormalisation Group And Gravity \rm\hfill\break
D.I.A.S. Preprint (1992)
\smallskip

\item{[\the\Zwiebach]} D.~Kutasov, Phys. Lett. \bf B220\rm, (1989), 153
\hfil\break
H.~Sonoda, Nucl. Phys.\bf B383\rm, (1992), 173\hfil\break
T.~Kugo and B.~Zwiebach, Prog. Theor. Phys.
\bf 87\rm, (1992), 801\hfil\break
K.~Ranganathan, {\it Nearby CFT's In The Operator
Formalism: The Role Of A Connection}, MIT pre-print MIT-CTP-2154
(1992)\hfil\break
K.~Ranganathan, H.~Sonoda and B.~Zwiebach,
\it Connections On The State
Space Over Conformal Field Theories\rm, MIT pre-print MIT-CTP-2193 (1993)
\smallskip

\item{[\the\WilKog]} K.G. Wilson and J. Kogut, Phys. Rep. \bf 12C\rm,
(1974), 75
\smallskip

\item{[\the\nlsr]} G.M.~Shore, Nucl. Phys, \bf B362\rm, (1991), 85
\smallskip

\item{[\the\HE]} S.W.~Hawking and G.F.R.~Ellis,   {\sl The Large Scale
Structure Of Space-Time \rm C.U.P. (1973)
\smallskip

\item{[\the\gx]} I.T.~Drummond and G.M.~Shore, Phys. Rev \bf D19\rm,
(1979), 1134\hfil\break
H.~Osborn, Phys. Lett. \bf B222\rm, (1989), 97\hfil\break
I.~Jack and H.~Osborn, Nucl. Phys. \bf 343\rm, (1990), 647\hfil\break
G.M.~Shore, Nucl. Phys. \bf B362\rm, (1991), 85
\smallskip

\item{[\the\lassig]} M. L\"assig, Nucl. Phys. \bf B334\rm,
(1990), 652
\vfill\eject\end